\documentclass[12pt,preprint]{aastex}
\def\msun{\rm M_{\sun}}

\def\LOri{${\lambda}$ Orionis}
\def\av{${\rm A_V}$}

\begin{document}
\shortauthors{Hernandez et al.}
\shorttitle{ Debris disks in the \LOri cluster}

\title{ {\em Spitzer} Observations of the $\lambda$ Orionis cluster I: the frequency of young debris disks at 5 Myr}

\author{Jes\'{u}s Hern\'andez\altaffilmark{1}, Nuria Calvet \altaffilmark{2}, 
L. Hartmann\altaffilmark{2}, J. Muzerolle\altaffilmark{3}, R. Gutermuth \altaffilmark{4,5}, 
 J. Stauffer\altaffilmark{6}}

\altaffiltext{1}{Centro de Investigaciones de Astronom\'{\i}a, Apdo. Postal 264, M\'{e}rida 5101-A, Venezuela.}
\altaffiltext{2}{Department of Astronomy, University of Michigan, 830 Dennison Building, 500 Church Street, Ann Arbor, MI 48109, US}
\altaffiltext{3}{Space Telescope Science Institute, 3700 San Martin Dr., Baltimore, MD 21218, US}
\altaffiltext{4}{Five Colleges Astronomy Department, Smith College, Northampton, MA  01027}
\altaffiltext{5}{Department of Astronomy, University of Massachusetts, Amherst, MA  01003}
\altaffiltext{6}{{\em Spitzer} Science Center, Caltech M/S 220-6, 1200 East California Boulevard, Pasadena, CA 91125}

\email{jesush@cida.ve}

\begin{abstract}
We present IRAC/MIPS {\em Spitzer} observations of intermediate-mass stars in the 5 Myr old {\LOri} cluster. 
In a representative sample of stars earlier than F5 (29 stars), we find a population of 9 stars 
with a varying degree of moderate 24{\micron} excess comparable to those produced by 
debris disks in older stellar groups. As expected in debris disks systems, those stars 
do not exhibit emission lines in their optical spectra. We also include in our study the
star HD 245185, a known Herbig Ae object which displays excesses in all {\em Spitzer} bands 
and shows emission lines in its spectrum. 
We compare the disk population in the {\LOri} cluster with 
the disk census in other stellar groups studied using similar methods 
to detect and characterize their disks and spanning a range of ages 
from 3 Myr to 10 Myr. We find that for stellar groups of 
5 Myr or older the observed disk frequency in intermediate mass stars (with spectral types from late B to early F) 
is higher than in low mass stars (with spectral types K and M). 
This is in contradiction with the observed
trend for primordial disks evolution, in which stars with higher 
stellar masses dissipate 
their primordial disks faster. 
At 3 Myr the observed disk frequency in intermediate mass stars 
is still lower than for low mass stars indicating that second generation dusty disks
start to dominate the disk population at 5 Myr for intermediate mass stars. 
This result agrees with recent models of evolution of solids in the region 
of the disk where icy objects form ($>30$ AU), which  suggest 
that at 5-10 Myr collisions start to produce large amount of 
dust during the transition from runaway to oligarchic growth 
(reaching  sizes of $\sim$500 km)
and then dust production peaks at 10-30 Myr, when objects reach their maximum sizes ($\ge$1000 km). 

\end{abstract}


\section{Introduction}
\label{sec:int}

Planets are thought to form in circumstellar disks of dust and gas
that result from the collapse of rotating molecular cloud cores during the
star forming process. 
These
primordial optically thick disks evolve 
due to viscous processes by which the disk is
accreting material onto the star 
while expanding to conserve angular momentum
\citep{hartmann98}. 
In addition, dust grains in the disk are expected to grow in size
and settle toward the mid-plane of the disk \citep{weidenschilling97},
where planetesimals and planets can be formed \citep{backman93,kenyon09}.
The dispersal of primordial disks operate less efficiently 
for stars with lower masses \citep{muzerolle03, aurora05, lada06, carpenter06, hernandez07a, kennedy09}.
Particularly, for low mass stars (K5 or later), 90\% of the stars lose 
their primordial disks by about 5-7 Myr \citep[e.g.][]{haisch01,hernandez08},
while for objects in the mass range of Herbig Ae/Be (HAeBe; types B, A or F early) stars,
the corresponding time scale for primordial disk dissipation is less
than 3 Myr \citep{hernandez05}.

The HAeBe stars, which have primordial optically thick disks, are the precursors
of intermediate mass stars with debris disks (like Vega or $\beta$ Pic).
Since the estimated lifetime for circumstellar dust grains
due to radiation pressure, sublimation, and  Poynting-Robertson drag
is significantly smaller than the estimated ages for the stellar
systems, the gas-poor disks observed in Vega type objects must be
replenished from a reservoir, such as sublimation of comets or collisions
between parent bodies \citep{chen06, kenyon04,dominik03}.
Most observational studies of the evolution of debris disks have focused
on the long time scales \citep[e.g][]{decin03, rieke05, su06, siegler07, carpenter09}.
These studies show that the luminosities of the debris disks 
decay following a simple power law (with characteristic time-scales of several hundred
millions years), as the debris reservoir of collisional parent bodies diminishes
and the dust is removed \citep{kenyon04,dominik03}. 
Observations suggest a peak in the debris disk phenomenon in 
intermediate mass stars at 10-30 Myr \citep{hernandez06,currie08a}; 
at the peak, debris disks are more frequent  and have larger infrared 
excesses. These observational results are supported by theoretical 
models of evolution of solids in the disk \citep{kenyon04,kenyon08}.
Since debris disks may trace active planet formation \citep{greenberg78,backman93,kenyon08,wyatt08,cieza08}, 
studies of debris disks younger than 
10-30 Myr are crucial to understanding the processes in which primordial disks evolve 
to planetary systems.

The {\LOri} cluster (also known as Collinder 69) represents a 
unique laboratory for studies of disk evolution because 
it is reasonably near and relatively populous, making  possible statistically
significant studies of disk properties in a wide range of stellar masses.
Based on main sequence fitting of early-type stars, \citet{murdin77}
obtained a distance of 400$\pm$40 pc and age of 4 Myr. Similarly,
using Stromgren photometry and main sequence fitting, \citet{dolan01} 
derived a distance of 450$\pm$50 pc and a turnoff age of 6 Myr.
Recently, \citet{mayne08} reported that the {\LOri} cluster and 
the $\sigma$ Orionis cluster have an age of 3 Myr. However, 
their disk frequencies \citep{barrado07,hernandez07a} and 
comparison between empirical isochrones \citep{hernandez08} suggest 
that the {\LOri} cluster is older than the $\sigma$ Orionis cluster.
For consistency with previous studies of disk and stellar population,
we assume a distance of 450 pc and an age of 5 Myr 
\citep[see][]{mathieu08}.

The unprecedented sensitivity and spatial resolution provided by
the {\em Spitzer Space Telescope} in the near- and mid-infrared
windows are powerful tools to expand significantly our understanding
of star and planet formation processes.
In this contribution, we analyze the near- and mid-infrared  properties 
of intermediate mass stars in the $\lambda$ Orionis cluster. 
 This paper is organized as follows. In \S \ref{sec:obs} we describe 
the observational data and the selection of intermediate mass star 
in the cluster. In \S \ref{disk_type} we present the method for identifying 
stars with infrared excesses. In \S \ref{disk_comp} we compare 
the disk population of the {\LOri} cluster  with results 
from other stellar groups inferring the nature of disk emissions 
around intermediate mass stars. Finally, we give our conclusions in \S \ref{sec:conc}.

\section{Observations}
\label{sec:obs}

\subsection{{\em Spitzer} observations}
\label{sec:ir}
We have obtained near-infrared and mid-infrared data of the 
{\LOri} cluster using the InfraRed Array Camera\citep[IRAC,][]{fazio04} 
with its four photometric channels (3.6, 4.5, 5.8 \& 8.0 \micron) and the 
24{\micron} band of the Multiband Imaging Spectrometer for {\em Spitzer} 
\citep[MIPS][]{rieke04} on board the {\em Spitzer Space Telescope}. These 
data were collected in mapping mode during 2004 October 11 (IRAC) and 
March 15 (MIPS) as part of G. Fazio's Guaranteed Time Observations 
in {\em Spitzer} program 37. 

The IRAC observations were done using a standard raster map with 280" offsets,
to provide maximum areal coverage while still allowing $\sim$20'' overlap
between frames in order to facilitate accurate mosaicking of the data.
Images were obtained in high dynamic range (HDR) mode, in which a short 
integration (1 s) is immediately followed by a long integration (26.8 s).
Standard Basic Calibrated Data (BCD) products from version S14.0.0 of
the {\em Spitzer} Science Center's IRAC pipeline were used to make
the final mosaics. Post-BCD data treatment was performed using
custom IDL software \citep{gutermuth04} that includes modules for 
detection and correction of bright source artifacts, detection and removal 
of cosmic ray hits, construction of the long and short exposure HDR mosaics, 
and the merger of those mosaics to yield the final science images with 
a scale of 1\arcsec.22 pixel$^{-1}$. Since mosaics of channels 4.5  and 8.0 
{\micron} have a 6.5{\arcmin} displacement to the north from the mosaics 
of channels 3.6 and 5.8 {\micron}, the region with the complete IRAC data 
set (57.6\arcmin x54.6\arcmin; hereafter IRAC region) is smaller than 
the field of view (FOV) of individual mosaics (57.6\arcmin x61.2\arcmin).
IRAC point source photometry was obtained using  the {\it apphot} package 
in IRAF, with an aperture radius of 3\arcsec.7 and a background annulus 
from 3.7 to 8\arcsec.6.  We adopted zero-point magnitudes for the standard 
aperture radius (12\arcsec) and background annulus (12-22\arcsec.4) of 19.665, 
18.928, 16.847 and 17.391 in the [3.6], [4.5], [5.8] and [8.0] bands, 
respectively. Aperture corrections were made using the values described in 
IRAC Data Handbook \citep{reach06}. Final photometric errors include the 
uncertainties in the zero-point magnitudes ($\sim$0.02 mag).

MIPS observations were made using medium scan mode with full-array
cross-scan overlap, resulting in a total effective exposure time per pointing
of 40 seconds.  MIPS observations cover more than 99\% of the IRAC region.
The images were processed using the MIPS instrument team Data
Analysis Tool (DAT), which calibrates the data and applies a distortion 
correction to each individual exposure before combining it into a final mosaic 
\citep{gordon05}. A second flat correction was also done to each image using 
a median of all the images in order to correct for dark latents and scattered 
light background gradients. The second flat correction is determined by 
creating a median image of all the data in a given scan leg, and then dividing 
each individual image in that scan leg by that median. Bright sources and 
extended regions are masked out of the data before creation of the median, 
so the nebulosity seen in the map has a negligible effect on any noise
that might be added by the second flat.  We obtained point source photometry
at 24 {\micron} with IRAF/{\it daophot} point spread function fitting,
using an aperture size of about 5\arcsec.7 and an aperture correction
factor of 1.73 derived from the STinyTim PSF model. The aperture size
corresponds to the location of the first airy dark ring, and was chosen
to minimize effects of source crowding and background inhomogeneities.
The absolute flux calibration uncertainty is $\sim$4\% \citep{engelbracht07}.
The photometric uncertainties are dominated by the background/photon noise.
Our final 24 {\micron} flux measurements are complete down to about 0.8mJy.

\subsection{Sample selection and optical spectra}
\label{s:sel}

Our procedure for identifying the disk population around 
intermediate mass stars in the {\LOri} cluster is similar
to that described in \citet{hernandez06}.
To find the bright stellar population of the {\LOri}
cluster, we selected from the Two Micron All Sky Survey (2MASS) catalog 
\footnote{Vizier Online Data Catalog, 2246 (R.M. Cutri et al. 2003)}
stars with J$\le$11. This limit was calculated using the V magnitude 
\citep{cox00} and the V-J color \citep{kh95} for a main sequence star 
with spectral type F5 located at 450 pc \citep{dolan01,mathieu08} and 
assuming a visual extinction of \av$\sim$0.4 \citep{diplas94}. 
The reddening in the 2MASS bands was calculated from the assumed visual extinction 
using the \citet{ccm89} extinction law with $R_V$=3.1. Our preliminary 
sample includes 159 stars located in the IRAC region. 

We found two relatively bright stars (V$<$12) without 2MASS counterpart 
when
comparing our preliminary sample with photometric studies of the bright 
population in the {\LOri} cluster \citep{dolan01,murdin77, duerr82}. 
One star (HD36862) is  a close companion ($\le$3\arcsec) of the central 
bright star of the cluster (the {\LOri} star), not resolved in the 
IRAC/MIPS images. 
The other star, HD245168, was included previously as member of the 
{\LOri} cluster with spectral type B9 \citep{dolan01,murdin77}.
Visual inspection of the 2MASS images reveals that HD 245168 is a 
stellar source not included in the 2MASS catalog. We estimated 2MASS 
photometry for this source using differential photometry. We use 
2MASS sources located inside 6{\arcmin} of radius to get the median value 
of  differential photometry between the 2MASS magnitudes and the 
instrumental magnitudes obtained on the 2MASS images using an aperture 
radius of 3{\arcsec} and a background annulus from 3{\arcsec} to 7{\arcsec}. 
Our estimation for HD 245168 indicates 2MASS photometry of J=9.50, H=9.52 and 
K=9.45 with a photometric error of 5\%; we added this star to our preliminary sample. 
Using the \citet{kharchenko01} catalog (limiting magnitude of V=12-14)  
and visual inspection on the IRAC 3.6{\micron} image, we did not find 
additional bright stars not included in the 2MASS catalog.

\begin{figure}
\epsscale{0.8}
\plotone{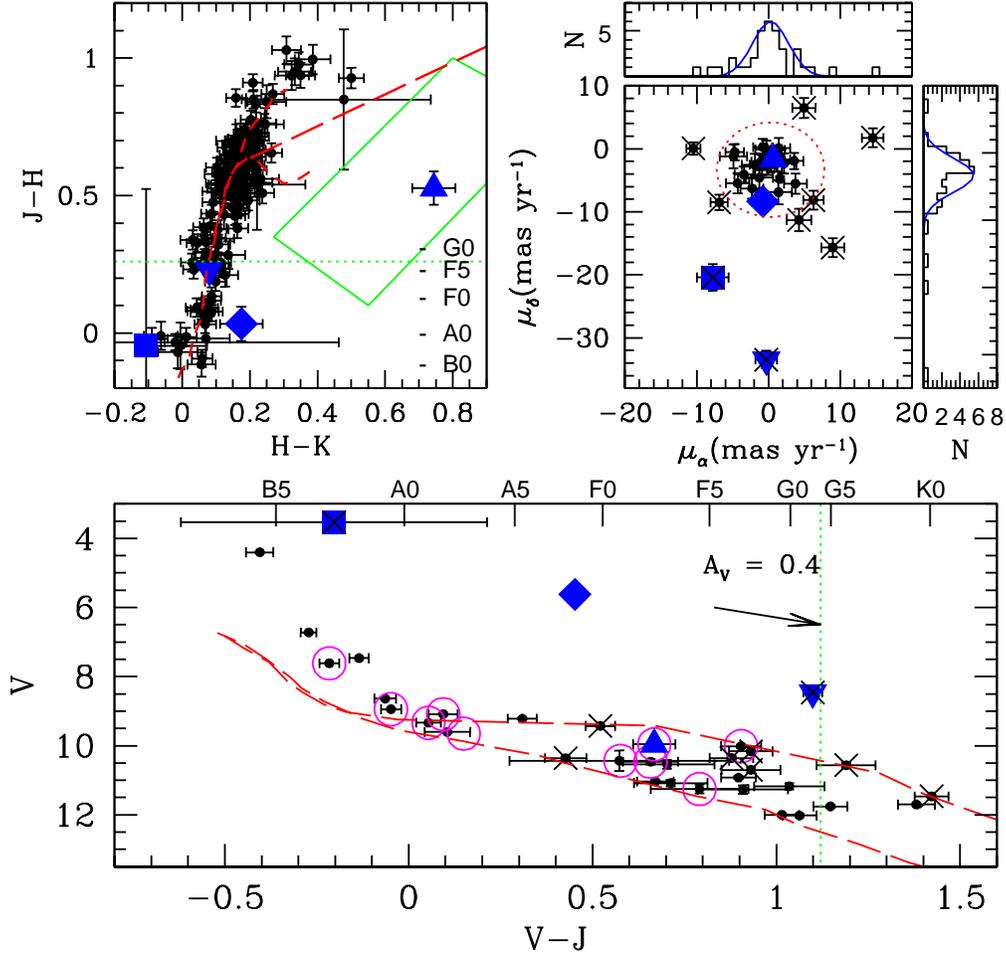}
\caption{Diagrams illustrating the selection of photometric 
candidates of the {\LOri} cluster. The HAeBe star HD 245185 (triangle), 
the star Lambda Orionis (square), the stars HD 36881 (diamond) 
and HD 37148 (inverse triangle) are marked.
The upper-left panel shows the 2MASS color-color diagram. The standard sequences 
from \citet{bessell88} are shown in dashed lines. 
The locus of the CTTS \citep{meyer97} and the HAeBe stars \citep{hernandez05} are defined by 
a long-dashed line and by a solid box, respectively. The photometric limit for a F5 star 
with reddening of \av=0.4 is represented by a dotted line in the 
upper-left 
panel and in the bottom panel.
We selected stars below the photometric limit.
The upper-right panel shows the vector point diagram for stars selected in the upper-left panel. 
In general, these stars have similar proper motions;
however, since the supernovae explosion 
could affect the kinematic properties of the cluster, kinematic criteria can not be used alone
to reject probable member of the cluster. Stars with discrepant proper motion are plotted with 
crosses. Finally, the bottom panel shows the color-magnitude diagram V-J versus V.  Our sample 
of early type star in the {\LOri} cluster are stars located 
left-ward (bottom panel) the photometric limit. 
We rejected as member the stars HD 36881 and HD 37148 which are located above the expected sequence.
Large open circles represent stars with infrared excesses 
(see \S \ref{disk_type}.) } 
\label{f:sel}
\end{figure}

Figure \ref{f:sel} illustrates the procedure to select early type 
candidates (F5 or earlier) of the {\LOri} cluster. The NIR
color-color diagram (upper-left panel) for the preliminary sample   
reveals that most of the stars are located on the main-sequence locus 
\citep{bessell88} indicating that, in general, these stars have little
extinction and no NIR excesses. Only two stars appear to have large 
extinction or NIR excesses. One star (HD 245185; V1271 Ori), a well known
Herbig Ae star with spectral type A5 \citep[e.g.,][]{wade07,acke05,
hernandez04,winter01,herbst99, finkenzeller84}, appears in the HAeBe star loci 
\citep{hernandez05}, where the NIR excess can be explained
by emission from an optically thick primordial disk with a sharp dust-gas
transition at the dust destruction radius \citep{dullemond04}.  The other 
star (HD 36881) is a spectroscopic binary with a peculiar spectral type 
B9III \citep{pedoussaut88}. 
Since HD36881  does not lie near the ZAMS 
(even allowing for photometric errors or binarity), probably this star 
is a foreground nonmember \citep[][ see also lower panel of Figure 
\ref{f:sel}]{dolan01}. The dotted horizontal line in the NIR color-color 
diagram is the [J-H] limit corresponding to a F5 star with \av=0.4.
We identified 35 early type candidates located below 
this limit. We also included the  Herbig Ae (HAe) star HD245185 as 
a early type star of the {\LOri} cluster. 

The upper-right panel shows the vector point diagram for the early 
type candidates. Proper motions are from \citet{roser08}. In this panel, 
we also show distributions of proper motions which can be 
represented by Gaussians centered at $\mu_{\alpha}*cos(\delta)\sim0.0$ 
mas yr$^{-1}$ and $\mu_{\delta}\sim0.3$ mas yr$^{-1}$, with $\sigma$=2.5 
mas yr$^{-1}$.  Most of the stars are located in a well defined region 
represented for the 3$\sigma$ criteria circle in the vector point plot;
however, the proper motion of the star {\LOri} does 
not agree with the general trend of the cluster. To explain these peculiar 
proper motions (as other peculiar kinematic properties), several authors 
\citep[e.g.,][]{maddalena87,cuhna96, dolan99} have invoked the presence 
of a massive companion of the {\LOri} star that became a supernova (SN), affecting 
the {\LOri}'s kinematics and removing  nearby molecular gas at the center
of the cluster about 1 Myr ago when the supernova exploded \citep{dolan01}. 
If this scenario is correct, then the SN explosion may have affected the
kinematics of other members of the cluster
specially if those members were part of a
multiple system, {\LOri} + SN, or were still in formation and embedded 
in their molecular cloud. So we can not use kinematic properties 
alone as a reliable criteria for membership.

The lower panel in Figure \ref{f:sel} shows the color magnitude diagram, V versus V-J, 
for the early type candidates. Most stars are located near or in the region defined by the 
ZAMS and the 5 Myr isochrone \citep{sf00}. Since 
it is well known that theoretical isochrones 
not including the birthline
do not accurately match empirical isochrones for 
intermediate mass stars \citep{hartmann03}, and theoretical evolutionary models 
often ignore some hydrodynamic processes in stellar models \citep{mamajek08},
these theoretical isochrones are plotted as reference only and are not used 
for our selection of the photometric candidates. However, comparing empirical 
isochrones in the low mass star range, we showed that the {\LOri} cluster is 
in similar evolutionary stage as other groups with ages quoted as 5 Myr \citep{hernandez08}, 
which is in agreement with previous results \citep[e.g.,][]{barrado07,dolan02}.
Two stars (HD36881 and HD37148) are located above the general trend described 
by the sample. The star HD36881 was reported previously as a foreground star 
\citep{dolan01} as discussed above, and HD37148 has the largest proper 
motion of the sample ($\mu_\delta$=-33.5 mas yr$^{-1}$; see upper-right panel), 
even larger than the proper motion of the star {\LOri}.
We excluded those two stars from being members of the cluster. In this contribution, 
we also excluded stars located right-ward from the dotted line, which represents
the limit corresponding to a F5 star with \av=0.4. Our final sample includes
29 early type candidates of the {\LOri} cluster.

\begin{figure}
\epsscale{1.0}
\plotone{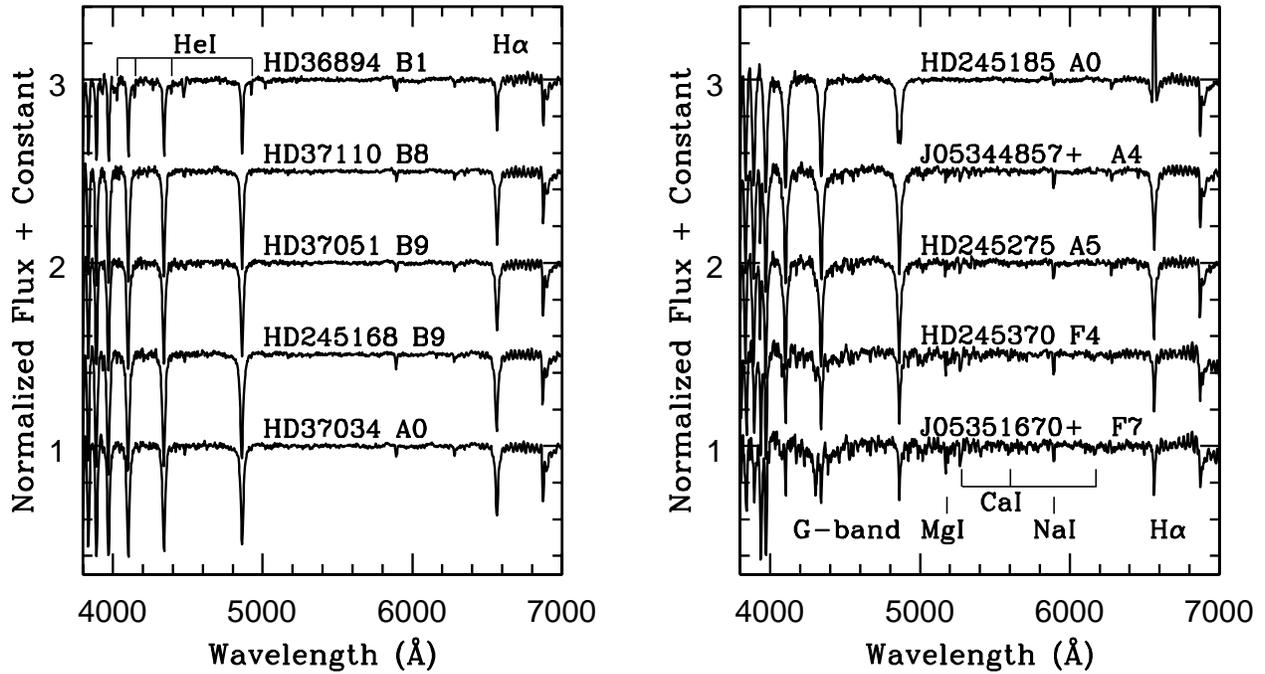}
\caption{Optical spectra of the intermediate mass stars bearing disks sorted by spectral types. 
The spectra were obtained with the CCDS spectrograph on the 1.3m Telescope 
of the MDM observatory. Spectral types were obtained using the SPTCLASS code, some features
used for spectral classification are marked in the panels. Emission lines are detected 
only in the HAe star HD 245185 }
\label{f:spectra}
\end{figure}

Stars with detected infrared excesses (see \S\ref{disk_type}) 
are encircled in the lower panel of Figure \ref{f:sel}. Optical spectra were 
obtained for these stars using the 1.3m McGraw-Hill telescope of the MDM 
Observatory equipped with the Boller \& Chivens CCD Spectrograph (CCDS). We 
used the 150 grooves per mm grating centered at 5300\AA. This configuration
provides a nominal resolution of $\sim$5{\AA} with spectral coverage 
from 3800{\AA} to 7100{\AA}. In Figure \ref{f:spectra} we show spectra for 
the stars with infrared excesses sorted by spectral types. Spectral
types were obtained using the SPTCLASS tool \footnote{http://www.astro.lsa.umich.edu/$\sim$hernandj/SPTclass/sptclass.html}, 
an IRAF/IDL code based on the methods described in \citet{hernandez04}.
The HAe star HD 245185 shows emission lines suggesting
that magnetospheric accretion processes are present in this star \citep{muzerolle04}.
The other stars do not show emission lines in their optical spectra indicating 
that magnetospheric accretion has stopped; this characteristic is typically observed in 
diskless stars and stars bearing debris disks. Most stars not observed 
with the CCDS have spectral types from \citet{hdc93}. 
For the stars without spectral type information, we estimated the latest 
spectral type they may have by
assuming no reddening and 
interpolating the observed V-J color in the table of standard colors given by \citet{kh95}.

Table \ref{t:mem} shows optical and {\em Spitzer} data for the early type candidates. 
Columns (1) and (2) show the name and the 2MASS denomination of the stars. Column(3), (4), (5) and (6)
show the IRAC magnitudes. Column (7) shows the 24 {\micron} MIPS fluxes. The typical 
errors in the IRAC and MIPS magnitudes are 0.02-0.05 and 0.03-0.06, respectively.   
Column (9) shows the excess ratio of the measured flux at 24{\micron} from that expected
from the stellar photosphere (see \S \ref{disk_type}). Spectral types and references are in 
columns (10) and (11). 
 Visual magnitudes and references are in columns (12) and (13). 
The last column shows the disk type around each star (see \S \ref{disk_type} and \S \ref{disk_comp}).

\section{Disk diagnostic}
\label{disk_type}

\subsection{Identifying stars with infrared excess}
\label{excess}
\begin{figure}
\epsscale{0.8}
\plotone{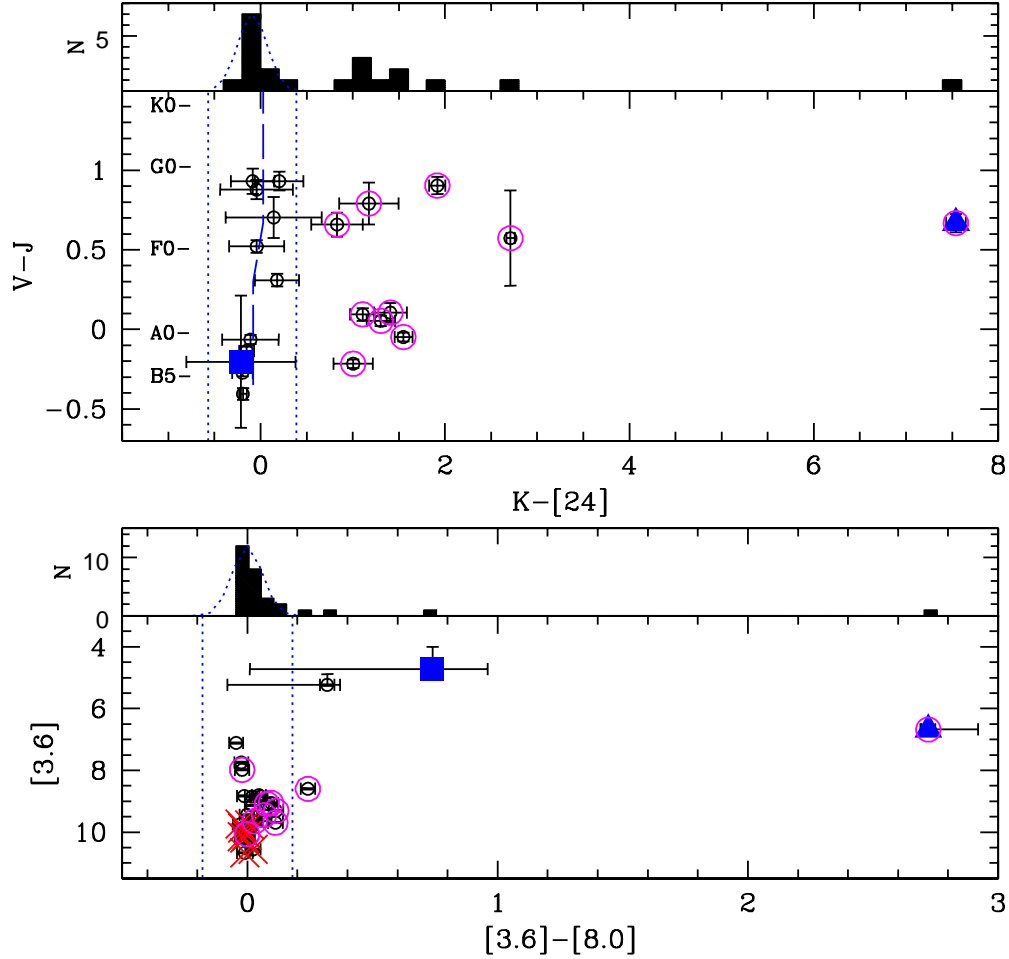}
\caption{ V-J vs. K-[24] color-color diagram for intermediate mass stars  
in the {\LOri} cluster (Upper Panel). Dotted lines define the locus expected
for diskless stars based on the K-[24] color distribution. Nine stars exhibit 
moderate excess at 24{\micron} (K-[24]$<$3) while the HAe star (HD 245185; triangle) 
exhibits strong excess at 24{\micron} (K-[24]$\sim$7.5). The bottom panel shows 
the color-magnitude diagram,  
[3.6] vs. [3.6]-[8.0],
used to detect stars with 
excess at 8\micron. Dotted lines define the photospheric limit based 
on the [3-6]-[8.0] color distribution. 
The two brightest stars in this plot are affected by saturation at 3.6{\micron}. Since these stars 
dot not show excess at 24\micron, the excess detected at 8{\micron} is not real. 
Thus, in addition to the HAe star (triangle), 
we have detected one star with small excess at 8{\micron} (HD 245370).  The square represents 
the central star {\LOri}}
\label{f:disks}
\end{figure}

To characterize the disk population among stars of the {\LOri} cluster, 
we need to identify photometric candidates that exhibit excess emission
at the IRAC/MIPS bands. Since a disk produces greater excess emission 
above the stellar photosphere at longer wavelengths, we selected the 
bands at 8{\micron} and 24{\micron} to identify stars with infrared 
emission due to the presence of disks. First, we identified stars 
with 24{\micron} infrared excess above the photospheric level 
in the upper panel of Figure \ref{f:disks}.
In the upper portion of this panel, we plot 
the K-[24] color distribution for the sample. The distribution of stars
around the expected photospheric color (K-[24]$\sim$0) 
describes a Gaussian centered at K-[24]=-0.09 with $\sigma$=0.16.
The 3$\sigma$ boundaries (vertical dotted lines) represent the photospheric 
colors; thus stars with excess at 24{\micron} have colors K-[24]$>$0.4.
Besides the HAe star, which displays the largest excess at 24{\micron} in 
our sample, we detected 9 stars (hereafter debris disk candidates) 
with a varying degree of 24{\micron} excesses comparable to the debris 
disk population in other young stellar regions \citep[e.g.][]
{young04,hernandez06,gorlova07,currie08a,currie08b},
and with no emission lines in their optical spectra (see Figure \ref{f:spectra}).  
As reference, we displayed the location of the spectral type sequence, using the standard
V-J colors from \citet{kh95}, and the corresponding K-[24] photospheric 
colors from the  STAR-PET tool of {\em Spitzer Science Center}
\footnote{http://ssc.spitzer.caltech.edu/tools/starpet/}

The lower panel of Figure \ref{f:disks} shows the [3.6]-[8.0] color 
- [3.6] magnitude diagram illustrating the procedure to detect 
infrared excess at 8{\micron}. The photospheric limits are defined 
using a 3$\sigma$ criteria based on the [3.6]-[8.0] color distribution, 
which can be represented by a Gaussian centered at [3.6]-[8.0]$\sim$0 
with $\sigma$=0.06. The error bars include the saturation effect, that
were estimated using the deviation from the expected photospheric colors
for the brightest objects detected in each IRAC band. Thus, 
the apparent 8{\micron} excess observed in stars with [3.6]$<$6 mag
are produced by saturation effect in the [3.6] band.   
Besides the HAe star which displays excesses in all IRAC bands, 
the star HD 245370  shows small excess at 8{\micron}; this star 
also shows the third largest excess at 24{\micron} ( K-[24]$\sim$ 2).

\subsection{Spectral Energy Distributions}
\label{seds}

\begin{figure}
\epsscale{0.8}
\plotone{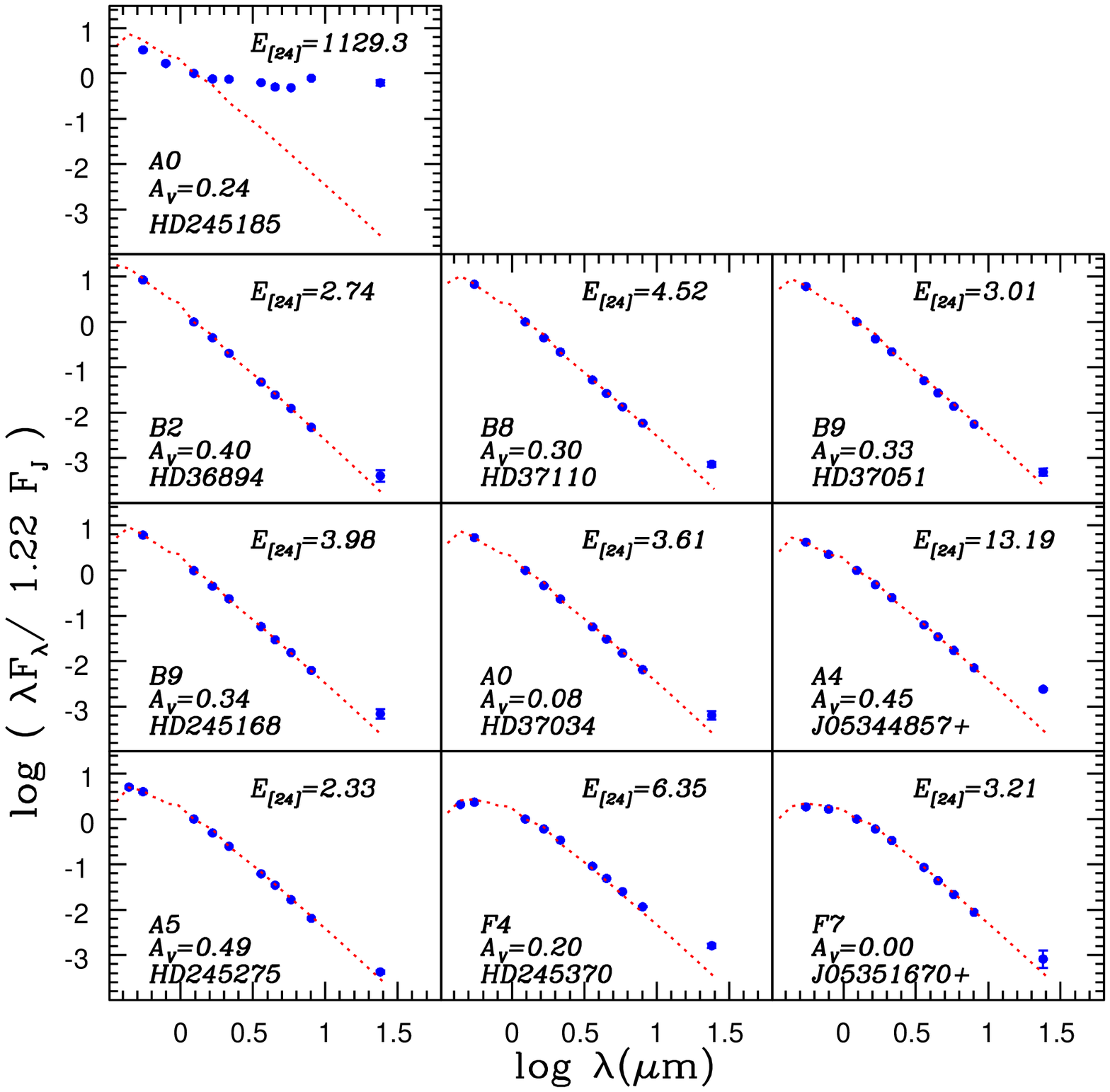}
\caption{SEDs for intermediate mass stars with infrared excess. Each panel shows the spectral type
of the object (see \S \ref{s:sel}), and the excess ratio, E24 \citep{rieke05},
calculated from the K-[24] color. Dotted lines show the corresponding photospheric colors \citep{kh95}.
Except for the star HD 245185, \av's were calculated using the V-J color. Since the J magnitude of HD 245185 
can be contaminated by disk emission, we have used the color B-V to calculate the reddening in this HAe star.
All SEDs are corrected by reddening and normalized at the J-band.}
\label{f:sed}
\end{figure}

We have plotted the SEDs for the disk bearing stars found in our 
early type sample of the {\LOri} cluster. Figure \ref{f:sed} shows the SEDs for 
the Herbig Ae star HD 245185 and for the debris disk candidates 
sorted by spectral types. 
The color excess,  E$_{V-J}$, was calculated 
using the observed V-J color and the standard color obtained interpolating
the spectral type (see \S \ref{s:sel}) in the calibration given by \citet{kh95}.
Using the E$_{V-J}$ and the extinction relation for normal interstellar 
reddening \citep{ccm89}, we obtained the visual extinctions reported in 
Figure \ref{f:sed}. Since the J magnitude of the HAe star could have contribution 
from the disk, we calculate the {\av} for this star using the B-V color 
from \citet{kharchenko01}.  
Out of 10 stars plotted in Figure \ref{f:sed},  
two stars (20\%) have {\av} larger than the value used as cut-off in \S \ref{s:sel} 
({\av}=0.4). Since stars bearing disks are expected to have larger extinctions
produced by the circumstellar material, we can estimate that the completeness 
in our selection of stars earlier than F5 is larger than  80\%. A F0 star needs 
to have {\av}$>$0.8 to be rejected from our sample (see Figure \ref{f:sel}).

Dotted lines in these plots  represent the standard colors 
(or photospheric levels) for a given spectral type \citep{kh95}.
Figure \ref{f:sed} also shows the excess ratio at 24{\micron} \citep{rieke05},
which was calculated  using the relation: $E_{24} = 10 ^{0.4*(K-[24] + 0.09)}$, 
where -0.09 is the median value of K-[24] for the 
photometric candidates with MIPS detections (see \S \ref{excess}).
In summary, we detected 10 early type disk-bearing stars
in the {\LOri} cluster. The HAe star HD 245185 is the only star that 
shows infrared excesses in all the {\em Spitzer} bands, in agreement with 
the presence of a primordial optically thick disk; emission lines 
in its optical spectrum indicates that HD 245185 is accreting material 
from the disk to the star.  Only the debris disk candidate HD 245370 shows 
infrared excess at 8\micron, the remaining debris disk candidates 
only exhibit infrared excess at 24\micron.

\section{Debris disk frequency in young stellar populations}
\label{disk_comp}

In \citet{hernandez06}, we reported that debris disks in 
the range of ages from 2.5 to 150 Myr are more frequent and 
have larger 24{\micron} excess around 10 Myr.   
This peak in the debris disk phenomenon is in agreement with 
models of evolution of solids in the region of the protoplanetary 
disk where icy planets are predicted to form 
\citep[30-150 AU;][]{kenyon04,kenyon05,kenyon08}. 
Observational results from the double cluster h and {$\chi$} Persei 
\citep[10-15 Myr;][]{currie08a} and the cluster NGC2232 
\citep[25 Myr;][]{currie08b} support our finding.  
However, it is still an open question whether the disk 
emission observed at 24{\micron} in debris disks candidates younger 
than 10 Myr originates from remaining primordial dust 
or from second generation dust produced by collisional 
cascades in the disk \citep[e.g.][]{hernandez06, hillenbrand08, wyatt08, cieza08}.  
To tackle  this question, we have compared the disk population in the 
{$\lambda$} Orionis cluster to disk populations of several 
stellar groups located within 500 pc of the Sun detected and characterized with the same methods 
\citep[][Hern\'{a}ndez et al, in preparation]{hernandez06,hernandez07a,hernandez07b,hernandez08}.

\begin{figure}
\epsscale{0.8}
\plotone{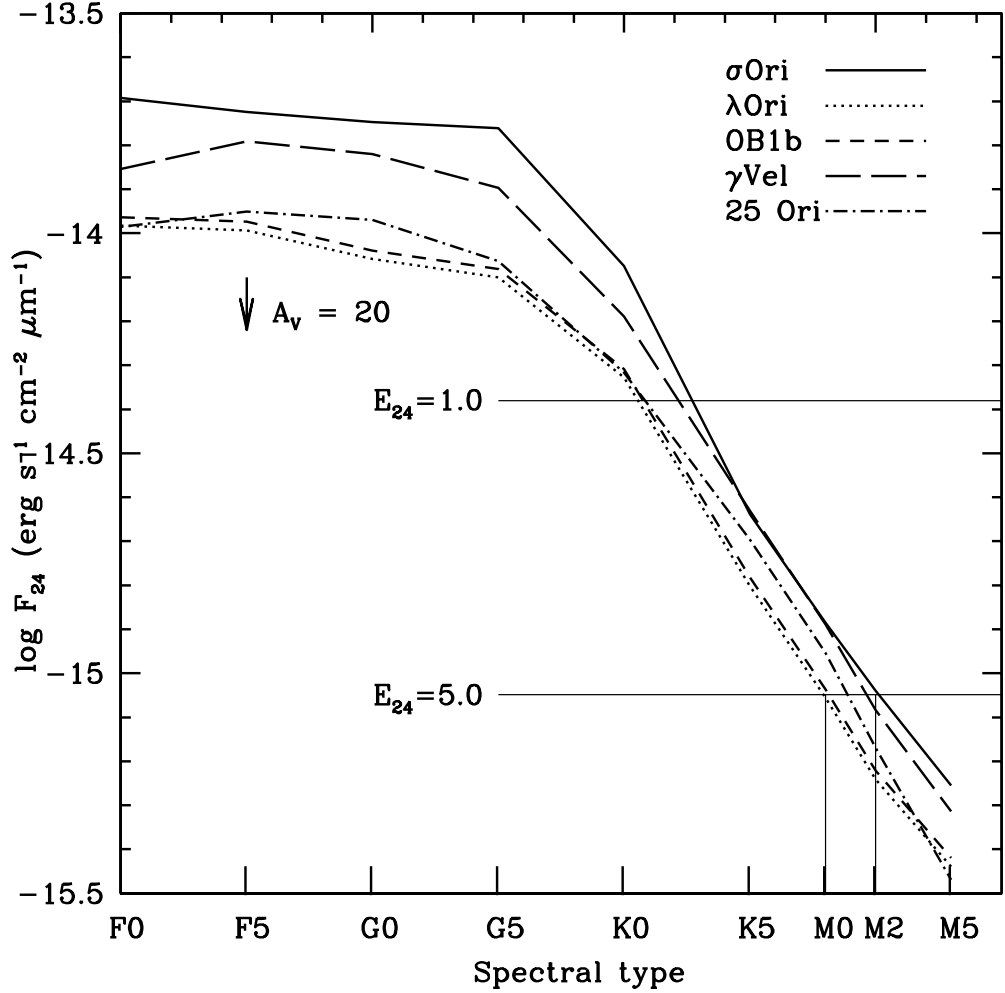}
\caption{Theoretical stellar fluxes at 24{\micron}. Combining  the \citet{sf00} evolutionary tracks 
and the corresponding K-[24] photospheric colors from STAR-PET tool of {\em Spitzer Science Center}
we estimated the theoretical flux expected for the stellar populations with distances, ages and 
extinctions listed in Table \ref{t:clusters}. Given the limiting flux at 24 {\micron} (0.6-0.7 mJy),
we plotted the limit where we can detect any excess at 24{\micron} (E$_{24}$ =1). For reference, 
we also plotted the limit where we can detect stars bearing disks with E$_{24} \ge$ 5. Vertical thin lines
define the spectral type range in which we can detect stars bearing disks with E$_{24} \ge$ 5. 
The Arrow represents reddening vector of \av=20. }
\label{f:det}
\end{figure}

Figure \ref{f:det} shows the theoretical stellar 
fluxes at 24{\micron} for diskless stars in the stellar 
populations of Table \ref{t:clusters}.  Using the \citet{sf00} 
evolutionary tracks and the properties listed in Table \ref{t:clusters},
we estimated the apparent K magnitude for stars in each stellar group. 
The 24 {\micron} fluxes were calculated using the corresponding K-[24] 
photospheric colors obtained from the STAR-PET tool of {\em Spitzer Space Center} 
and the zero point magnitude defined for MIPS observations  
($\sim$ 7.17 Jy \footnote{http://ssc.spitzer.caltech.edu/mips/calib/}). 
Given the limiting flux at 24 {\micron} (0.5-0.7 mJy), and using the 
\citet{sf00} evolutionary tracks to estimate the K magnitude for stars 
in each stellar group, we inferred that we can detect any excess at 24 {\micron} 
in stars F0 or earlier with low or moderate reddening. Figure 
\ref{f:det} shows the limit for diskless stars (E$_{24}$=1).
The detection thresholds at 24{\micron} for intermediate mass stars in different 
clusters are given by the width of the photospheric region defined 
by the K-[24] color distributions. 
These detection thresholds are nearly the 
same in all stellar populations of Table \ref{t:clusters}, in which we can detect 24{\micron} 
excess in intermediate mass stars with E$_{24} > 1.6$. 
We also show in Figure \ref{f:det} 
the detection threshold for stars bearing disks with E$_{24}$=5.
Figure \ref{f:det} indicates that for all our surveys the spectral 
type limit for detecting stars bearing disks with E$_{24} \ge$5 
is M1$\pm$1.   

Using the same method for 8 {\micron}, we estimated that a 
diskless star in our surveys with spectral type M5 has [8.0]$<$13. 
This is well above of the IRAC detection limits of our surveys ([8.0]$\sim$14.0-14.5). 
The detection threshold at 8{\micron} for a M5 star is defined by the error of the IRAC SED slope. 
The IRAC SED slope, used to identify infrared excess at 8{\micron} 
in low mass stars,  is calculated from the [3.6]-[8.0] color and its 
error is dominated by the photometric error at [8.0]. Thus, using the 
typical error in the IRAC SED slope at [8.0]=13 ($\sim$ 0.25 for all surveys), 
we can detect disks around stars with spectral type M5 or earlier 
with a excess ratio at 8 {\micron} larger than 1.2.

\begin{figure}
\epsscale{0.8}
\plotone{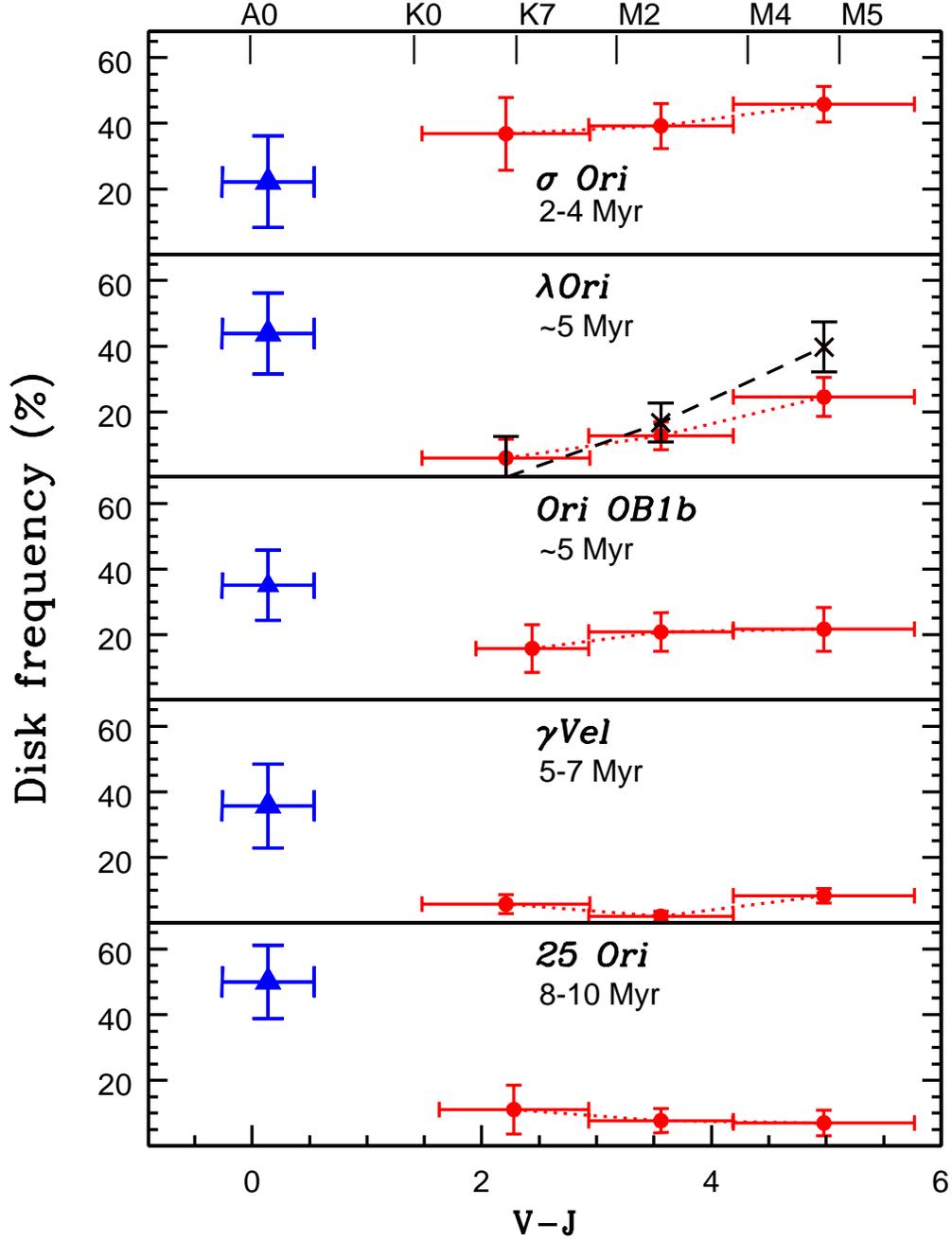}
\caption{Disk frequency versus spectral types for several star forming regions.
In the stellar population of 3 Myr, disk frequencies for A-type 
stars are lower than disk frequencies around low mass stars in agreement 
with the expected trend for primordial disk dissipation. In contrast, at 5 Myr 
and older, disks around A-type stars are more frequent than in low mass stars 
indicating than second generation dusty disk must dominated the disk population
at higher stellar masses. The low mass bins are centered at the spectral types 
K6.5, M2.5 and M4.5. Using these values we calculated a reference 
detection threshold for each bin (see Figure \ref{f:det}): 
$E_{24}\sim$ 3, 7, 10 (from left to right). The detection threshold for the
intermediate mass bin are given by the width of the photospheric region 
in each stellar group (e.g. Figure \ref{f:disks}), $E_{24}\sim$1.6. 
For the {\LOri} cluster, we plotted with X's the 
disk frequencies  calculated from the disk census in \citet{barrado07}. 
Since the IRAC thresholds defined in \citet{barrado07} do not include 
photometric errors, the number of disk bearing  stars in the last bins 
could be overestimated.}
\label{f:disk_frec}
\end{figure}

Figure \ref{f:disk_frec} shows 
observed 
disk frequencies 
as a function of color/spectral type for 
several young stellar populations, the clusters
$\sigma$ Ori, {\LOri}, $\gamma$ Velorum, a region
in the Orion OB1b subassociation, and the stellar
aggregate 25 Ori. Frequencies were calculated for several 
bins of spectral types in each stellar population.

The bin corresponding to the intermediate mass stars includes 
stars ranging from B8 to F0; most of the intermediate mass stars 
with infrared excesses are in this range \citep{hernandez06,hernandez07a,hernandez08}.
In general, the intermediate mass stars of the stellar populations plotted in Figure \ref{f:disk_frec} 
have spectral type information available in the literature \citep{houk78,houk99, 
nesterov95, kharchenko01,caballero07,hernandez05,hernandez06}; 
optical photometric colors were used to estimate spectral types for
stars without spectral type information. In agreement with the extinctions
listed in Table \ref{t:clusters}, the reddening for intermediate mass stars
in these stellar groups are relatively low. 
Except for the star HD36219 located in the Orion OB1b subassociation (with \av=1.46),  all 
stars in the intermediate mass samples plotted in Figure \ref{f:disk_frec} have \av$<$1.

Disks around low mass stars were split in three bins of spectral types:
from K0.5 to M0.5, from M0.5 to M3.5 and from M3.5 to M6 (hereafter; the low mass bins).
Given the low reddening expected for the stellar populations
plotted in Figure \ref{f:disk_frec}, photometric colors represent a good 
approximation of spectral types when spectroscopic information is not 
available for the stellar groups. Thus, we used the standard table of 
colors given by \citet{kh95} to obtain spectral type estimations.
Observed disk frequencies of low mass 
stars in the {\LOri} cluster were calculated analyzing the IRAC and MIPS photometry of members confirmed 
spectroscopically by \citet{dolan01, barrado07, maxted08}, and \citet{sacco08} using 
radial velocity distributions and the presence of \ion{Li}{1} in absorption (Hernandez et al in preparation). 
{\em Spitzer} point source photometry for these low mass members 
was obtained using the same methods described in \S \ref{sec:ir}.
Disk detection follow the method described in \citet{hernandez07a, hernandez07b,hernandez08}.
We used NIR photometry from the 2MASS catalog \citep{cutri03} 
and optical photometry from \citet{dolan02} and \citet{barrado07}.
Since the R-J color is more complete for the spectroscopic members compiled for the {\LOri} cluster,
we used this color to calculate photometric spectral types and separate the sample
in the bins described above. 
For comparison, we plotted the disk frequencies obtained from the disk 
census in \citet{barrado07}. Since the IRAC photospheric region 
defined in \citet{barrado07} does not take into account photometric 
errors, the number of stars bearing disks could be overestimated in
the lowest mass bins ( Maria Morales Calderon, private communication).

For the $\sigma$ Orionis cluster, we used the disk census in \citet{hernandez07a}. To reduce the 
contamination by non-members, observed disk frequencies of low mass stars were calculated using the spectroscopic 
members compiled in \citet{hernandez07a}, members confirmed by \citet{sacco08} and \citet{maxted08}, 
and the X ray members from \citet{franciosini06}. Since most of this sample does not have spectral types,
we used the color V-J in \citet{hernandez07a} and the standard colors \citep{kh95} to
separate the low mass sample in the low mass bins.

For the Orion OB1b subassociation  and the 25 Orionis aggregate we used the 
disk census in \citet{hernandez07b}. Observed disk frequencies were calculated
using confirmed members and spectral types calculated using our 
SPTCLASS code \citep[][Brice\~{n}o et al in preparation]{briceno05,briceno07}. 
Since our samples of confirmed members in the Orion OB1b 
subassociation and the 25 Orionis do not go as early as K0.5, the first low mass star bin in these groups 
does not cover the same spectral type range as the other stellar populations in Figure \ref{f:disk_frec}
(from K2 for the 25 Orionis aggregate and from K4 for the Orion OB1b subassociation).

Finally, for the $\gamma$ Velorum cluster we used the disk census in \citet{hernandez08}. 
Observed disk frequencies were calculated using the number of members expected 
as a function of V-J color \citep{hernandez08};
this color was used to separate the sample in the low mass bins.
Using empirical Li-depletion studies and empirical isochrone comparisons, 
the $\gamma$ Velorum cluster appears to be slightly older than the {\LOri} cluster 
\citep[5-7 Myr;][]{jeffries09} and younger than the 25 Orionis 
stellar aggregate \citep{hernandez08}.

In Figure \ref{f:disk_frec}, the observed disk frequencies in the low mass stars bins
were calculated using the infrared excess at 8 {\micron}. Since a disk produces
greater emission above the stellar photosphere at longer 
wavelengths, those stars with excess at 8 {\micron} are expected to have 
excess at 24 {\micron}. However, since the sensitivity of the 24 {\micron}
MIPS band is lower than that of the IRAC 8 {\micron} channel, some of the faintest 
stars bearing disks are below the MIPS detection limit. 
We also included as stars bearing disks, stars with 
small or no infrared excesses in the IRAC bands 
but detectable excess at 24 {\micron} (transitional disk candidates).
The transitional disk candidates represent a relative small fraction of 
the disk-bearing stars in these stellar groups 
\citep[$\lesssim$10\%;][]{hernandez07b,ercolano09,uzpen09}, 
so they do not significantly affect our conclusions.

The infrared excesses observed in low mass stars are unlikely from second generation dust.
{\em Spitzer} observations of 314 solar-type stars\footnote{ranging in spectral types from late F 
to early K and in the range of ages from 3 Myr to 3 Gyr} in the Formation and Evolution of 
Planetary Systems (FEPS) Legacy program \citep{carpenter09} 
indicate that the dust temperature of debris disks ($T_{dust}$) 
ranges  from $\sim$50 K to $\sim$200 K,
with a fractional dust luminosity ($f_{dust}=L_{dust}/L_*$) less than 10$^{-3}$. 
Adopting these values, following the two-parameter model of \citet{wyatt08},
 and using the stellar parameters for a K0 star at 3, 5 and 10 Myr \citep{sf00},
we estimated that debris disks around low mass stars show small 
or no excess at 8{\micron} (E$_8$ $<$ 1.2) and modest excess at 24{ \micron} ($E_{24}$ $<$ 4).
The two-parameter model includes two fundamental observable parameters of debris disks 
($T_{dust}$ and $f_{dust}$) and it is useful to estimate the expected infrared excesses 
produced by a second generation disk 
with a unique $T_{dust}$;
however, notice that we need a more detailed model to
reproduce the SED that for some disks can only be explained by the presence 
of several dust populations with different temperatures.
In general, low mass stars bearing disks in our census show excesses 
much larger than those expected for debris disks (E$_{24}>$25), 
thus the low mass bins in 
Figure \ref{f:disk_frec} represent frequencies of primordial disks with 
little or no contamination by second generation dust.

In contrast, the two-parameter model \citep{wyatt08}  applied to debris disks around 
A0 stars with dust temperatures ranging from 50 K to 200 K and  
fractional luminosity of 10$^{-3}$ indicates 
maximum $E_{24}$ values of 29, 21 and 21 
at ages of 3, 5, and 10 Myr, respectively. 
In general, debris disks around intermediate mass stars exhibit less infrared 
excess than these maximum levels \citep[e.g.][]{rieke05, hernandez06, currie08a,wyatt08}.
Studying the MIPS observations of 160 A-type stars (with age range
from 5 Myr to 850 Myr), \citet{su06} show that debris disk systems
have f$_{dust}<$10$^{-3}$ and most of them ($>$90\%)are characterized 
by dust temperatures between 50 K and 200K (although for stars older 
than 400 Myr the dust temperatures of the disks appear in a narrow 
range between 50 and 150 K).
Since our debris disks candidates have similar levels of 24{\micron}
excess than those reported in older debris disks systems 
\citep[e.g.,][]{rieke05,su06,gorlova07,currie08a},  we expect a
similar range of T$_{dust}$ and f$_{dust}$
Debris disks with $T_{dust} \ge$ 200 K  (and/or $f_{dust} \ge 10^{-3}$) are very 
scarce \citep{zuckerman04,williams06,su06,bryden06,rhee07,wyatt08,uzpen09}.
This kind of massive debris disks (like HR4796A and $\beta$ Pic) 
can be above our estimations of E$_{24}$ and could be explained 
if the observed dust is produced by eventual collisions between 
two large objects \citep[$>$ 1000 km;][]{kenyon04, kenyon05, song05}. 
However, an alternative explanation is that those objects are in 
a phase between HAeBe star and true debris systems \citep{rieke05} 
retaining some of the primordial dust. The star HD245370 shows 
excess at 8{\micron} that can be explained by the two-parameter 
model assuming a $T_{dust} \sim$ 250 K, however this star is outside 
of the intermediate mass bin plotted in  Figure \ref{f:disk_frec}.
In our samples of intermediate mass stars, only 
three stars bearing disks  show excess at 24{\micron} larger than the theoretical 
limits calculated above: HD 290543 in the Orion OB1b subassociation ($E_{24}$=537.5), HD 245185 in the $\lambda$ 
Orionis cluster ($E_{24}$=1129.3), and V346 Ori in the 25 Orionis aggregate ($E_{24}$=376.4).
These three stars also show H$\alpha$ in emission indicating the presence of 
an accreting primordial disk. 
The remaining intermediate mass stars with disks 
exhibit characteristics normally present in debris disk systems, 
they have $E_{24} <$12.5 and no emission lines in their optical spectra.

In summary, in the low mass bins the detection threshold is such that only
primordial disks can be detected. In contrast, both primordial disks and debris disks 
can be detected in the intermediate mass bin. We may take then the observed frequency of 
disks in the low mass bins in Figure \ref{f:disk_frec} as indicator of the 
frequency of primordial disks. 
Figure \ref{f:disk_frec} shows that
the observed disk frequencies in K and M stars decrease
as the stellar groups become older.
Previous studies have shown that in young stellar populations primordial
disks dissipate faster with higher stellar masses, for instance, in 
IC348 \citep[2-3 Myr;][]{lada06, currie09a}, NGC 2264 \citep[$\sim$3][]{sung09},
and $\sigma$ Ori in Figure \ref{f:disk_frec}. 
According to this trend, the frequency of primordial disks around intermediate 
mass stars should decrease with age in proportion to the 
observed decrease in the primordial disk frequency in the
low mass bins.  
However, Figure \ref{f:disk_frec} shows that the observed frequency
of disks in the  intermediate mass bin increases after $\sim$ 5 Myr.
Since in this bin both primordial and debris disks can be detected,
and since the primordial disk frequency is expected to decrease,
the increase in disk frequency points to an increase in the frequency of debris disks.
Assuming a binomial distribution for the errors of the disk frequencies, the 
significance of this increase ranges from 68\%( 1$\sigma$) to 95\%(2.0 $\sigma$).
The increase in frequency of debris disks has been reported
before by \citet{hernandez06} and \citet{currie08b} at $\sim$ 10-15 Myr.
The reversal in the observed disk frequency in the intermediate mass bin  
shown in Figure \ref{f:disk_frec} shows that this increase is already
present at much earlier ages, at $\sim$ 5 Myr.
Enough secondary dust must have been created 
by 5 Myr to make second generation dust detectable.
This second generation dust must have been replenished from a reservoir, 
such as collisions between parent bodies \citep[e.g.,][]{chen06, dominik03,kenyon04}.

\citet{kenyon08} described a suite of numerical calculations of planets 
growing from ensembles of icy planetesimals at 30-150 AU in disks around 1-3 $\msun$ stars.
These models predict that the 24{\micron} excess from second generation dust rises 
at stellar ages of 5-10 Myr (during the transition from runaway to oligarchic growth, reaching sizes of $\sim$500 km), 
then  peaks at 10-30 Myr (when the first objects reach their maximum sizes,$\ge$1000 km) and finally 
slowly declines (as the reservoir of small solids in the disk diminishes). 
The expected theoretical trend has been observed previously by \citet{hernandez06} and 
\citet{currie08a, currie09b}. However, Figure \ref{f:disk_frec} suggests the second 
generation nature of the excesses observed at 24 {\micron} in debris disk candidates 
of 5 Myr or older. Since evolutionary processes are expected to occur faster in the 
inner disk than the outer disks for primordial disks \citep{weidenschilling97,dahm07,lada06,aurora06} 
and for debris disks \citep{kenyon04,kenyon05,kenyon08}, there is a possibility that the debris 
disk candidates in our studies have an outer primordial disk component which does not 
contribute significantly to the emission observed at $\lambda \le$ 24 \micron. An example 
of this kind of object is AU Mic, a $\sim$ 12 old  M-type star with a collisional 
evolved debris region in the inner part of the  disk, and with outer regions composed 
by pristine material that is still part of the remnant primordial disk \citep{metchev05}.  
Additional observations at longer  wavelength or observation of gas component 
of the disk are needed to confirm the complete dissipation of the outer 
primordial disks in the young debris disks candidates. 
Finally, \citet{currie09a} suggest that debris disks can be present in the 2-3 Myr
IC 348 cluster. Using theoretical models of dust removal \citep{takeuchi01}, they estimate 
that the dust removal timescale is less than 1/20th of the median age for IC 348 stars, 
even if there is some residual gas in an optically thin disk. Thus, we can not reject 
the possibility that some of the disks detected around intermediate mass stars 
of the $\sigma$ Orionis cluster are second  generation disks. We need additional
studies to confirm the actual nature of the intermediate mass stars bearing disks 
detected in this cluster.

\section{Summary and Conclusions}
\label{sec:conc}

We have used the IRAC and MIPS instruments on board the
{\em Spitzer Space Telescope} to study the frequencies and properties
of disks around the intermediate mass population of the 5 Myr old 
$\lambda$ Orionis cluster. We have selected 29 intermediate 
mass stars ($\sim$F5 or earlier) using optical and 2MASS photometry.
One star, HD 245185, displays excesses in all IRAC band, strong excess 
at 24 {\micron} and emission line in its optical spectra, 
supporting previous results in which HD 245185 is a HAe star having 
an accreting primordial disk \citep{finkenzeller84, herbst99, winter01, hernandez04, acke05, wade07}. 
Additionally, we have identified 
nine stars as debris disks, which exhibit modest excess at 24{\micron} and 
very small or no excesses in the IRAC bands. The SEDs of our debris disks sample 
are comparable to those observed in debris disk populations of older stellar groups.  
As expected in stars with second generation disks, these stars do not show 
detectable accretion indicators in their optical spectra. 

We have combined observations of the {\LOri} cluster with those 
of other stellar groups of ages raging from 3 to 10 Myr located at $<$ 500 pc from the sun, 
studied using similar methods to detect and characterize their disks. 
Because of observational thresholds,
the observed disk frequencies observed in K and M stars 
reflect the primordial disk populations,
while the observed disk frequencies in intermediate
mass stars encompass both primordial and debris disks.
As expected in primordial disk evolution 
the observed frequencies in K and M stars are smaller in older stellar groups.
However, observed disk frequencies in intermediate mass stars, 
represented by a bin in spectral type from B8 to F0, show a behavior 
in sharp contrast with that expected for primordial disk evolution. 
In particular, the disk frequency in intermediate mass star 
increases from $\sim$20\% at $\sim$3 Myr, to $\sim$40\% at 5 Myr 
and to $\sim$50\% at $\sim$10 Myr.
Since the timescale for primordial disk dissipation is 
dependent on stellar mass, the expected frequency of primordial 
disks around intermediate mass stars at ages $\ge$ 5 Myr, 
as inferred from the corresponding frequency in low mass stars, 
is very low.
In stellar population of 5 Myr or older, the observed disk frequencies
in intermediate mass stars are larger than in low mass stars.
Thus, the increasing in the observed disk frequency 
at these ages indicates that these disks are dominated by second generation dust.
This result agrees with models of evolution of solids in the disks \citep{kenyon08},
in which at 5-10 Myr collisions start to produce copious amount 
of dust during the transition from runaway to oligarchic growth.
When oligarchs reach sizes $\sim$500 km, collisions between 1-10 km
objects produce debris instead of mergers and the leftover
planetesimals are ground to dust.

\acknowledgements
We thank Thayne Currie and Maria Morales Calderon for insightful communications,
and the anonymous referee for input that greatly improved this manuscript.
This work is based on observations made with the {\em Spitzer Space Telescope} (GO-1 0037), 
which is operated by the Jet Propulsion Laboratory, California Institute of Technology under
a contract with NASA. J.Hern\'{a}ndez and N. Calvet gratefully acknowledge support from 
the NASA Origins program and the {\em Spitzer} General Observer program.

\clearpage

\begin{deluxetable}{ccccccccccccc}
\rotate
\tabletypesize{\scriptsize}
\tablewidth{0pt}
\tablecaption{Early type stars of the $\lambda$ Orionis Cluster \label{t:mem}}
\tablehead{
\colhead{ID} & \colhead{2MASS} & \colhead{[3.6]} & \colhead{[4.5]} & \colhead{[5.8]} & \colhead{[8.0]} & \colhead{F$_{24\micron}$} & \colhead{Excess ratio} & \colhead{Spectral} & \colhead{Ref.} & \colhead{Vmag} & \colhead{Ref.} & \colhead{Disk} \\
\colhead{ } & \colhead{ }      & \colhead{(mag)}        & \colhead{(mag)}       & \colhead{(mag)}     & \colhead{(mag)} & \colhead{(mJy)}  & \colhead{} & \colhead{type }    & \colhead{ } &\colhead{ } & \colhead{ } & \colhead{type}\\
}
\startdata

HD36861 & 05350831+0956036 & 4.728 & 4.381 & 4.006 & 3.988 & 169.31 & 0.89 & O5 & 2 & 3.532 & 4 & No disk\\
HD36822 & 05344923+0929225 & 5.235 & 5.002 & 4.994 & 4.916 & 69.81 & 0.91 & B0 & 2 & 4.402 & 4 & No disk\\
HD36895 & 05351280+0936478 & 7.114 & 7.163 & 7.146 & 7.159 & 9.01 & 0.91 & B3 & 2 & 6.722 & 4 & No disk\\
HD245203 & 05351380+0941494 & 7.741 & 7.749 & 7.733 & 7.765 & 5.49 & 0.94 & B8 & 2 & 7.465 & 4 & No disk\\
HD36894 & 05351670+0946394 & 7.982 & 7.978 & 7.973 & 8.003 & 12.28 & 2.74 & B2 & 1 & 7.614 & 4 & Debris \\
HD37035 & 05355825+0931541 & 8.852 & 8.795 & 8.785 & 8.833 & 2.04 & 0.98 & B9 & 2 & 8.638 & 4 & No disk\\
HD37110 & 05362962+0937542 & 9.055 & 9.082 & 9.075 & 8.960 & 7.34 & 4.53 & B8 & 1 & 8.944 & 4 & Debris \\
HD37051 & 05360418+0949550 & 9.087 & 9.045 & 9.039 & 9.009 & 4.97 & 3.01 & B9 & 1 & 9.087 & 4 & Debris \\
HD245140 & 05345817+0956267 & 8.803 & 8.772 & 8.740 & 8.757 & 2.67 & 1.28 & B9 & 2 & 9.218 & 4 & No disk\\
HD37034 & 05355938+0942480 & 9.311 & 9.271 & 9.300 & 9.196 & 4.70 & 3.61 & A0 & 1 & 9.335 & 4 & Debris \\
HD37159 & 05365811+1016586 & 8.849 & 8.787 & 8.781 & 8.796 & 2.11 & 1.05 & A3 & 2 & 9.427 & 4 & No disk\\
HD245168 & \nodata          & 9.452 & 9.459 & 9.420 & 9.385 & 4.44 & 3.98 & B9 & 1 & 9.605 & 4 & Debris \\
HD245185 & 05350960+1001515 & 6.679 & 6.201 & 5.498 & 3.958 & 4702.13 & 1129.9 & A0 & 1 & 9.959 & 4 & Primordial \\
HD245370 & 05360940+1001254 & 8.599 & 8.557 & 8.543 & 8.357 & 14.27 & 6.35 & F4 & 1 & 10.014 & 4 & Debris \\
HD244927 & 05335042+1004211 & 8.828 & 8.829 & 8.841 & 8.841 & 2.44 & 1.31 & A7 & 2 & 10.163 & 4 & No disk\\
HD244907 & 05335115+0946421 & 9.139 & 9.108 & 9.172 & 9.120 & 1.53 & 1.04 & F8 & 2 & 10.357 & 4 & No disk\\
HD244908 & 05334712+0940261 & 9.760 & 9.772 & 9.787 & 9.751 & \nodata & \nodata & A2 & 2 & 10.361 & 4 & No disk\\
\nodata & 05344857+0930571 & 9.694 & 9.631 & 9.634 & 9.582 & 11.38 & 12.46 & A4 & 1 & 10.437 & 5 & Debris \\
HD245275 & 05354485+0955243 & 9.648 & 9.553 & 9.604 & 9.619 & 2.13 & 2.33 & A5 & 1 & 10.463 & 4 & Debris \\
HD245385 & 05361338+0959244 & 9.653 & 9.626 & 9.637 & 9.630 & 1.13 & 1.24 & A0 & 2 & 10.536 & 4 & No disk\\
\nodata & 05335032+0958185 & 9.439 & 9.447 & 9.451 & 9.441 & 1.14 & 1.00 & F7 & 3 & 10.699 & 4 & No disk\\
HD245386 & 05362132+0950414 & 9.745 & 9.720 & 9.716 & 9.774 & \nodata & \nodata & A2 & 2 & 10.924 & 5 & No disk\\
\nodata & 05345914+0933508 & 10.143 & 10.172 & 10.181 & 10.164 & \nodata & \nodata & F3 & 3 & 11.036 & 5 & No disk\\
299-3 & 05360529+1021271 & 10.190 & 10.185 & 10.196 & 10.201 & \nodata & \nodata & F3 & 3 & 11.082 & 4 & No disk\\
\nodata & 05365226+0929584 & 9.862 & 9.856 & 9.864 & 9.881 & \nodata & \nodata & F9 & 3 & 11.174 & 4 & No disk\\
\nodata & 05352468+1011452 & 10.070 & 10.085 & 10.106 & 10.071 & 1.98 & 3.21 & F7 & 1 & 11.254 & 4 & Debris \\
\nodata & 05334028+0948013 & 10.103 & 10.096 & 10.092 & 10.104 & \nodata & \nodata & F6 & 3 & 11.265 & 4 & No disk\\
h-star & 05350920+1002518 & 10.670 & 10.637 & 10.703 & 10.680 & \nodata & \nodata & F8 & 3 & 11.999 & 5 & No disk\\
\nodata & 05354220+1013447 & 10.560 & 10.538 & 10.540 & 10.537 & \nodata & \nodata & G0 & 3 & 12.022 & 5 & No disk\\
\enddata
\tablenotetext{~}{References: (1) \citet{hdc93}; (2) This work; (3) photometric spectral types; (4) \citet{kharchenko01};  (5) \citet{dolan02}}
\end{deluxetable}

\begin{deluxetable}{cccccccc}
\tabletypesize{\scriptsize}
\tablewidth{0pt}
\tablecaption{Properties of stellar populations \label{t:clusters}}
\tablehead{
\colhead{Name} & \colhead{Distance} & \colhead{Ref} & \colhead{\av} & \colhead{ref} & \colhead{age} & \colhead{Ref} & \colhead{E$_{24} \ge$ 5 (*)} \\
\colhead{ } & \colhead{ (pc) }      & \colhead{ } & \colhead{(mag)}       & \colhead{} & \colhead{(Myr)} & \colhead{}  & \colhead{SpT}\\
}
\startdata
$\sigma$ Ori  & 440  & 1,2      & 0.2 & 9,10  & 2-4 Myr  &  2 & M2 \\ 
$\lambda$ Ori & 450  & 3,4      & 0.4 & 11    & 5 Myr    &  4 & M0 \\
Ori OB1b      & 440  & 5        & 0.4 & 5     & 5 Myr    &  5 & M0 \\
$\gamma$ Vel  & 350  & 6,7,8    & 0.2 & 12,13 & 5-7 Myr  & 13  & M2 \\
Ori OB1b      & 330  & 5        & 0.3 & 5     & 8-10 Myr &  5 & M1 \\
\enddata
\tablenotetext{*}{The last column represents the later spectral type in which we can detect infrared excess at 24{\micron} with E$_{24}\ge$5 (see Figure \ref{f:det})}
\tablenotetext{~}{References: (1) \citet{brown98}; (2) \citet{sherry08}; (3) \citet{dolan01}; (4) \citet{mathieu08}; 
(5) \citet{briceno07}; (6) \citet{leeuwen07};(7) \citet{millour07}; (8) \citet{north07}; (9) \citet{brown94};  (10) \citet{bejar99}; 
(11) \citet{diplas94};(12) \citet{pozzo00}; (13) \citet{jeffries09} }
\end{deluxetable}

\clearpage

\end{document}